\documentclass{emulateapj}

\usepackage{verbatim}

\newcommand{\rsol}{R_\odot}
\newcommand{\msol}{M_\odot}
\newcommand{\lsol}{L_\odot}
\newcommand{\logg}{\rm{log}~g}
\newcommand{\cval}{1.79}
\newcommand{\alphaval}{-0.82}
\newcommand{\betaval}{1.92}

\shorttitle{Techniques for Finding Close-in, Low-mass Planets around Subgiants}
\shortauthors{Medina et al.}

\begin{document}
.

\title{Techniques for Finding Close-in, Low-mass Planets around Evolved Intermediate Mass Stars }

%\author[0000-0001-8726-3134]{A. A. Medina}
%\affil{Harvard-Smithsonian Center for Astrophysics\\
%60 Garden St. Cambridge, MA 01238, USA}

%\author{J. A. Johnson}\affil{Harvard-Smithsonian Center for Astrophysics\\
%60 Garden St. Cambridge, MA 01238, USA}
%\author{J. D. Eastman}\affil{Harvard-Smithsonian Center for Astrophysics\\
%60 Garden St. Cambridge, MA 01238, USA}
%\author{P. A. Cargile}\affil{Harvard-Smithsonian Center for Astrophysics\\
%60 Garden St. Cambridge, MA 01238, USA}

\author{A. A. Medina\altaffilmark{1}}
\affil{Harvard-Smithsonian Center for Astrophysics, 60 Garden St.,Cambridge, MA 01238}

\author{J. A. Johnson\altaffilmark{1}}
\author{J. D. Eastman\altaffilmark{1}}
\author{P. A. Cargile\altaffilmark{1}}

\begin{abstract}
Jupiter-mass planets with large semi-major axes ($a > 1.0$ AU) occur at a higher rate around evolved intermediate mass stars. There is a pronounced paucity of close-in ($a < 0.6$ AU), intermediate period ($5 < P < 100$ days), low-mass ($M_{\rm planet} < 0.7M_{\rm Jup} $) planets, known as the `Planet Desert'. Current radial velocity methods have yet to yield close-in, low-mass planets around these stars because the planetary signals could be hidden by the (5-10) m~s$^{-1}$ radial velocity variations caused by acoustic oscillations. We find that by implementing an observing strategy of taking three observations per night separated by an optimal $\Delta t$, which is a function of the oscillation periods and amplitudes, we can average over the stellar jitter and improve our sensitivity to low-mass planets. We find $\Delta t$ can be approximated using the stellar mass and radius given by the relationship $\Delta t = $\cval $(M/M_{\odot})^{\alphaval} ~(R/R_{\odot})^{\betaval}$. We test our proposed method by injecting planets into very well sampled data of a subgiant star, $\gamma$ Cep. We compare the fraction of planets recovered by our method to the fraction of planets recovered using current radial velocity observational strategies. We find that our method decreases the RMS of the stellar jitter due to acoustic oscillations by a factor of three over current single epoch observing strategies used for subgiant stars. Our observing strategy provides a means to test whether the Planet Desert extends to lower mass planets.   

\end{abstract}

\section{Introduction}

The majority of confirmed exoplanets discovered to date are around stars that are less massive than $M < 1.3\msol$. The wealth of planets discovered around solar and later-type stars provides information about planetary formation and demographics in this mass regime. However, knowledge of planetary formation and statistical properties of planets as a function of stellar mass is incomplete due to difficulties with detection methods concerning main sequence, intermediate mass (IM) main sequence A- and F-type stars ($M > 1.3 \msol$).
	 
F-type stars have large levels of astrophysical noise  (50 m~s$^{-1}$), known as stellar jitter, due in part to pulsations and inhomogeneous convection \citep{Saar1998}. The increased jitter levels make it difficult to detect planets with radial velocity (RV) semi-amplitudes below or comparable to this noise level. Stars with M $>$ 1.25$\msol$ lack the convective envelope necessary to power a magnetic dynamo leading to the absence of stellar magnetic field.  For the majority of main sequence IM stars, the star does not spin down via magnetic braking and thus spins rapidly throughout its main sequence lifetime. The rapid stellar rotation broadens the relatively few, in comparison to a solar mass star, absorption lines intermediate mass stars have making the measurement of small line shifts needed for precise radial velocities extremely difficult.
	 
Moving off the main sequence and targeting subgiants, the evolved counterparts of A- and F-type stars, provides a means to bypass these difficulties and search for planets around IM stars. The star expands during the subgiant phase of evolution, it cools and decreases its rotation speed producing the rich population of narrow absorption lines necessary for precise radial velocity studies.  
	
\begin{figure}[ht] %%%%%%% FIGURE 
\includegraphics[scale = 0.6]{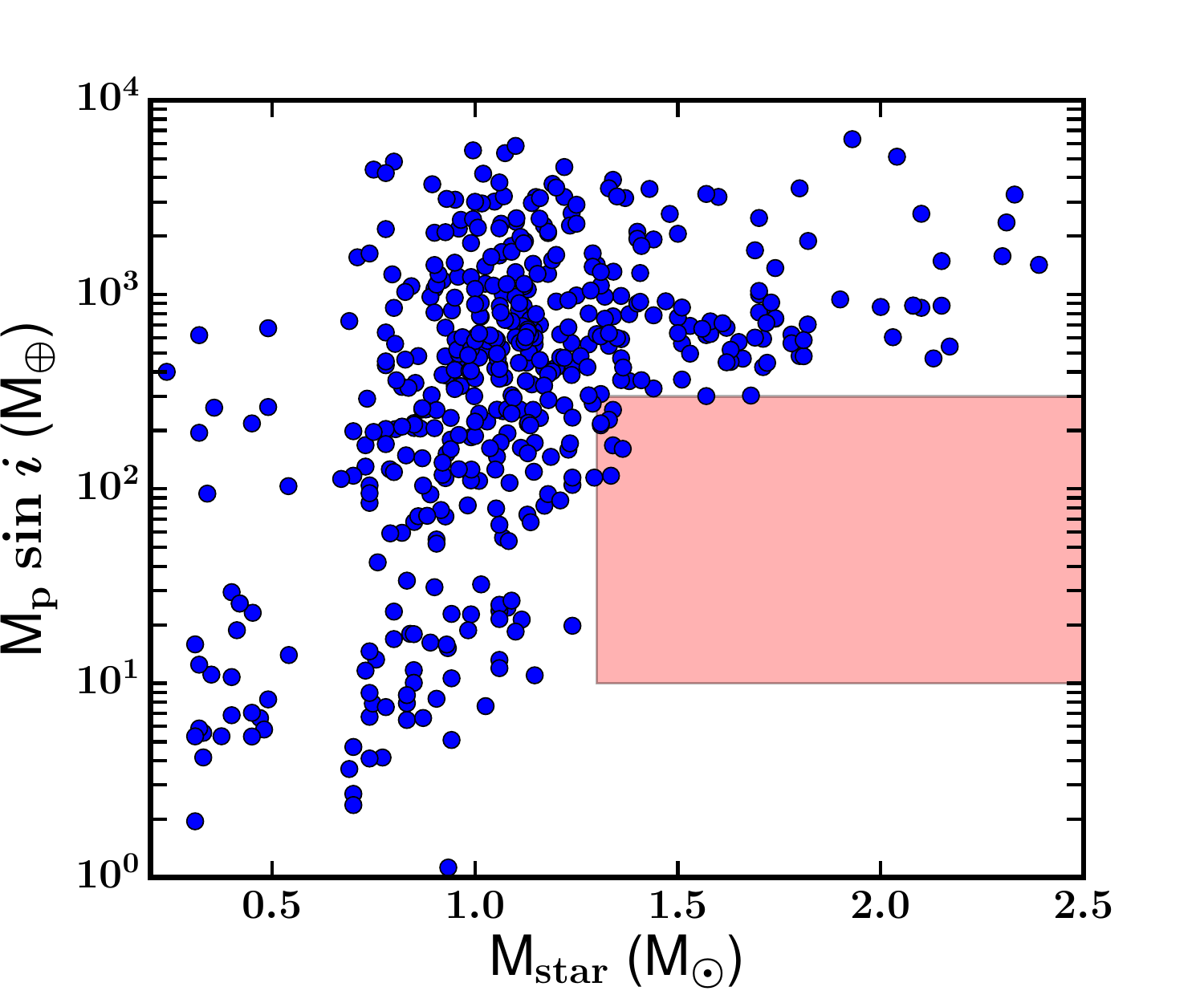}
\caption{Minimum mass of confirmed RV planets as a function of stellar mass (Exoplanets.org, 2/2/18). Shaded region is the target of this study.}
\label{fig:pd}
\end{figure}

Past surveys of subgiant stars have resulted in the discovery of many Jupiter and Super-Jupiter mass planets beyond 1 AU \citep{Hatzes2005,Johnson2007a, LovisMayor2007,Sato2008,Bowler2010,Reffert2015}, but also revealed a paucity of close-in ($a <$ 0.6 AU), intermediate period (5 $< P <$ 100 days), low-mass ($M_{\rm planet} <$ 0.7$M_{\rm Jup} $),  planets, known as the `Planet Desert' (Figure \ref{fig:pd}). Studies of these confirmed giant planets around subgiants have shown that giant planet occurrence increases as a function of stellar mass \citep{Johnson2010} and also have confirmed a definitive correlation between giant planet occurrence and stellar metallicity \citep{Johnson2010,Reffert2015} similar to the one found for main sequence stars \citep{Santos2004,Fischer2005,Buchhave2012,Dawson2013,Buchhave2015}.  
	
\begin{figure}[ht] %%%%%%% FIGURE 
\includegraphics[scale = 0.6]{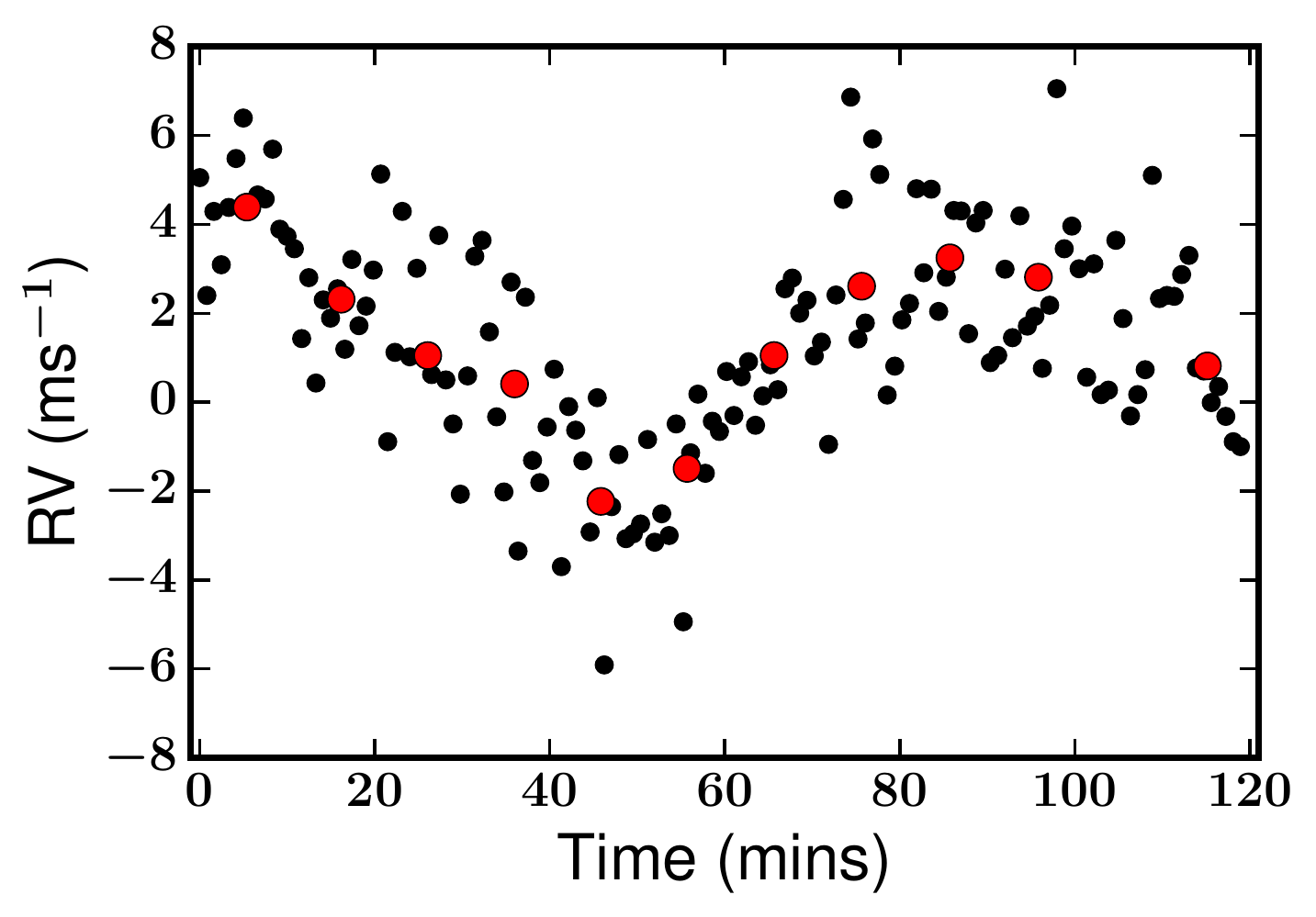}
\caption{High-cadence Radial Velocity data (black points) for HD142091 from Keck-HIRES. The red points are the data binned in 10 minute intervals}
\label{fig:pmodes_ex}
\end{figure}

Doppler tomography, an alternative detection technique that does not yield masses, has confirmed close-in $P <$ 5 days transiting planets around main sequence A- and F-type stars \citep{Rodriguez2017,Zhou2017,Stevens2017,Crouzet2017,Zhou2016}. Interestingly, \citet{Bowler2010} showed that the hot highly inflated Jupiter and Saturn-mass planets found around main sequence IM stars with Doppler tomography do not exist around their evolved counterparts, likely due to engulfment during stellar evolution, but the same cannot be said for the paucity of planets in the Planet Desert. These close-in, low-mass planets cannot be conclusively ruled out because past surveys have not been sensitive to the 5-10 m~s$^{-1}$ RV signal from a planet that is comparable to or smaller than the stellar jitter of an evolved IM star. An example of this jitter for subgiant HD142091 can be seen Figure \ref{fig:pmodes_ex}. These high-cadence data reveal the 5 m~s$^{-1}$ RV variations on the timescale of minutes that could potentially be reducing sensitivity of RV surveys to planets located in the Planet Desert.  

Acoustic pressure mode (p-mode) oscillations caused by turbulent convection near the stellar surface are the dominant source of stellar jitter (5-10 m~s$^{-1}$) in subgiant stars. \citet{KB1995}  further characterized the p-mode oscillations in the Sun. They found that the velocity amplitudes and p-mode oscillation frequencies in stars scale with stellar radius, mass, and effective temperature (T$_{\rm eff}$). These stellar properties are related to the physical cause of the oscillations: the convective flux, pressure scale height in the convection zone, and the stellar temperature. Thus, the timescale upon which p-mode oscillations occur as well as the amplitude of the oscillations, can be predicted. More recently, \citet{Dumusque2011} have shown RV variations, due to p-modes and granulation, could be averaged out with longer integration times and a higher frequency of intra-night observations, respectively. If the timescale upon which p-mode oscillations occur is known, it can be used to predict the optimum time between intra-night observations as suggested by \citet{Dumusque2011}. The dominant source of observed stellar jitter in subgiants can be modeled and thus reduced with the implementation of a proper observing strategy --- namely, taking three observations per night separated by a calculated optimal $\Delta t$ that is based on the mass, radius, effective temperature, and luminosity of the star. 
	
In this paper, we present observational techniques to search for planets in the Planet Desert. We use stellar properties to calculate the period and amplitude of the p-mode oscillations, then simulate their effect on the observed RV signal. The simulations determine the time between intra-night observations that most significantly reduces the observed dominant source of jitter in subgiants. We use ground based observations from the SONG telescope of a well known subgiant to test our proposed observing technique.  We find an empirical relationship between $\Delta t$ and the mass and radius of the star that can be used to inform observing strategies of future RV surveys.

\section{Methods}

Stars with convective envelopes exhibit p-mode oscillations, as the turbulent convective motion excites acoustic waves that ring in the normal modes of oscillation. These waves propagate through the star allowing for characterization of the stellar interior. Each mode is characterized by three integers; radial order $n$, which represents the number of nodes in the radial direction, angular degree $l$, the number surface nodes, and azimuthal order, $m$, the number of surface nodes that cross the equator. The azimuthal order only becomes important in stars with high rotational velocities, thus it will not be treated further for the purposes of our studies.

For high order $n$, low-degree $l$ systems that are characteristically observed in subgiant stars, the frequencies, $\nu$, of each mode can be approximated by Equation:  
\begin{equation}
\label{eq:freq}
\nu_{\rm nl} \simeq \Delta \nu \left(n + l/2 + \epsilon \right)
\end{equation}
\citep{Tassoul1980,Gough1996}.
\noindent Where $\Delta \nu$ is the large frequency separation and difference between consecutive radial nodes, $n$,  of the same angular degree $l$. $\epsilon$ is a constant based on stellar structure near the stellar surface. $\Delta \nu$ is related to the sound travel time through the center of the star and thus gives a good estimate of the mean stellar density:
\begin{equation}
\label{eq:density}
\Delta \nu \propto \left(M/R^3\right)^{1/2}
\end{equation}

The power spectrum created by the individual modes is composed of an envelope of frequencies with a characteristic $\nu_{\rm max}$, the frequency of maximum power, and $\Delta \nu$. In the Sun, $\nu_{\rm max}$, has a value of 3.05 mHz producing oscillations with periods of five minutes giving p-modes the name `solar-like oscillations'. These oscillations give rise to radial velocity variations, $v_{\rm osc}$, with amplitudes around 20 cm~s$^{-1}$ \citep{KB1995}. The amplitudes of these oscillations are proportional to the properties of the star. Larger, more luminous stars have larger amplitude oscillations in comparison to the oscillations in the Sun. Thus, the RV variations due to p-modes could make finding planets via the radial velocity method difficult.

\citet{Dumusque2011} describe observational methods to mitigate RV variations caused by p-modes and other sources of stellar noise, like granulation. They simulated radial velocity curves of stars ranging from spectral type G2 to K1 at different main sequence ages. They find that in comparison with the standard one observation per night strategy a well as other strategies  of most RV surveys, taking three 10 minute observations per night separated by two hours reduces the root-mean-square (RMS) radial velocity variation by 30 \%. They find that this observing cadence is able to average out the correlated red noise due to stellar granulation phenomena which occurs on timescales of hours to days better than observing once per night for 30 minutes or two 15 minute observations per night separated by two hours. They also find that p-mode RV variations can be averaged out with longer integration times.

The method proposed by \citet{Dumusque2011} treats all stars regardless of evolutionary state or spectral type equally, however \cite{KB1995} shows that the amplitudes as well as $\Delta \nu$ and $\nu_{\rm max}$ are functions of intrinsic stellar properties: mass, effective temperature, radius, and luminosity. \citet{KB1995} demonstrate that stellar properties can be used to predict $\Delta \nu$, $\nu_{\rm max}$, and the amplitudes of solar-like oscillations in other stars using Equations:

\begin{equation}
\label{eq:dnu}
\Delta \nu  = \left(M/ \msol \right)^{1/2}\,\left(R/\rsol \right)^{-3/2}\, 134.9 \,\mu \rm Hz
\end{equation}

\begin{equation} 
\label{eq:numax}
\nu_{\rm max} = \frac{M/\msol}{ \left(R/\rsol \right)^{2} \sqrt{T_{\rm eff}/5777\, K}}\, 3.05 \,\rm mHz
\end{equation}

\begin{equation}
\label{eq:amp}
A_{\rm vel} =  23.4\, {\rm cm~s^{-1}}\frac{L/\lsol~\left(\tau_{\rm osc}/\tau_{\rm osc_{\odot}}\right)^{0.5}}{\left(M/\msol \right)^{1.5} \left(T_{\rm eff}/5777 \right)^{2.25}} 
\end{equation} 

\noindent Where $\tau_{\rm osc}$ is the lifetime of the p-mode oscillation. 

In comparison to the Sun, $\nu_{\rm max}$ of a subgiant shifts to lower frequencies, thus p-modes have longer periods, and the star exhibits larger radial velocity amplitudes. For the subgiant star $\beta$ Gem (1.73$\msol$, 3.4$\lsol$, 8.21$\rsol$, T$_{\rm eff}$ = 4935 K), \citet{Stello2017} find $\nu_{\rm max}$ = 89 $\mu$Hz, with oscillation periods around 187 minutes, and $v_{\rm osc}$ = 5 m~s$^{-1}$ \citep{Hatzes2012}. The p-mode oscillation spectrum for a subgiant has vastly different properties than for a Sun-like star. In particular when comparing the periods of oscillations, Sun-like stars exhibit periods on the order of minutes, which could be averaged over in a typical (10-30) minute exposure of a star \citep{Dumusque2011}, while oscillations of subgiants occur on the timescales of hours, making it difficult and an inefficient use of telescope time to average out these oscillations in one exposure. 

In this section we describe an observational technique used to average out p-mode radial velocity variations in subgiants. We present an observing strategy that incorporates multiple observations in one night following \cite{Dumusque2011}, but with a calculated optimal $\Delta t$ that is motivated by the characteristic p-mode RV variations of the star predicted from the work of \citet{KB1995,Mosser2013,KB2011}. 

\subsection{Simulated p-mode Radial Velocity Curves}
To calculate the optimal $\Delta t$ for a given star, we simulated the p-mode radial velocity variations based on the mass, radius, luminosity, effective temperature of the star. In order to simulate RV variations due to p-modes, we must determine the frequencies and amplitudes of oscillations. To predict the frequencies of the p-mode oscillation spectrum,  we used Equation \ref{eq:freq}. The variables $\Delta \nu$,  $\epsilon$, and $n$ are determined using Equation \ref{eq:dnu}, and Equations: 

\begin{equation}
\label{eq:epsilon}
\epsilon = \nu_{\rm max}/ \Delta \nu - n_{\rm max}
\end{equation}

\begin{equation}
\label{eq:nmax}
n_{\rm max} \simeq 22.6 \left( \frac{M/\msol}{(T_{\rm eff}/5777K)(R/\rsol)}\right)^{1/2}- 1.6
\end{equation}

For the oscillation mode radial order, $n$, in Equation \ref{eq:freq}, we used $n_{\rm max} $ and $n_{\rm max}~\pm$ 1,2,3 because this provides the envelope of frequencies that is characteristically observed in p-mode power spectra.  Because observations of a star are from light integrated over the stellar disk, due to geometric cancellation, only $l = 0,1, 2, 3 $ are observed \citep{KB1995}. However, because the amplitudes of the oscillations decrease with increasing angular degree, $l = 3$ is only observed under optimal observing conditions. We wanted to characterize the dominant radial velocity variations with the largest semi-amplitudes, so we chose to use $l  = 0,1$ in Equation \ref{eq:freq}. For reference, \citet{Jimenez1990} measured p-mode velocity amplitudes in the Sun. They find for radial order $n$ = 19, the amplitude for $l$ = 0 to be 19~cm~s$^{-1}$, and $l$ = 1 to be 29~cm~s$^{-1}$. The semi-amplitudes of oscillations corresponding to each frequency were predicted from a Gaussian envelope centered on $\nu_{\rm max}$ with maximum amplitude calculated using Equation \ref{eq:amp} with a full-width at half maximum equal to 7$\Delta \nu$ adapted from a study on simulating p-mode oscillations of the giant star $\xi$ Hyi by \citet{Stello2004}. Although $\xi$ Hyi is a giant, whose oscillation spectrum may have a slightly different FWHM, because we are only using $n_{\rm max}~\pm 0, 1,2,3$, the innermost, regions of the envelope, the full width of the envelope will not affect our simulations. An attempt by \citet{Chaplin2009} to find an empirical relationship for the prediction of oscillation lifetimes. However, \citet{Baudin2011} claim because of the large uncertainty of ground based observations and inhomogeneous sample that was used in the \citet{Chaplin2009} study, their empirical relationships derived for $\tau_{\rm osc}$ may be incorrect.  Therefore, oscillation lifetimes were estimated from a normal distribution centered around 2.5 days with a standard deviation of 1.5 days following the results from \citet{Chaplin2009} in a study of mode lifetimes of various well known subgiants.

\begin{figure}[ht] %%%%%%% FIGURE
\includegraphics[scale = 0.6]{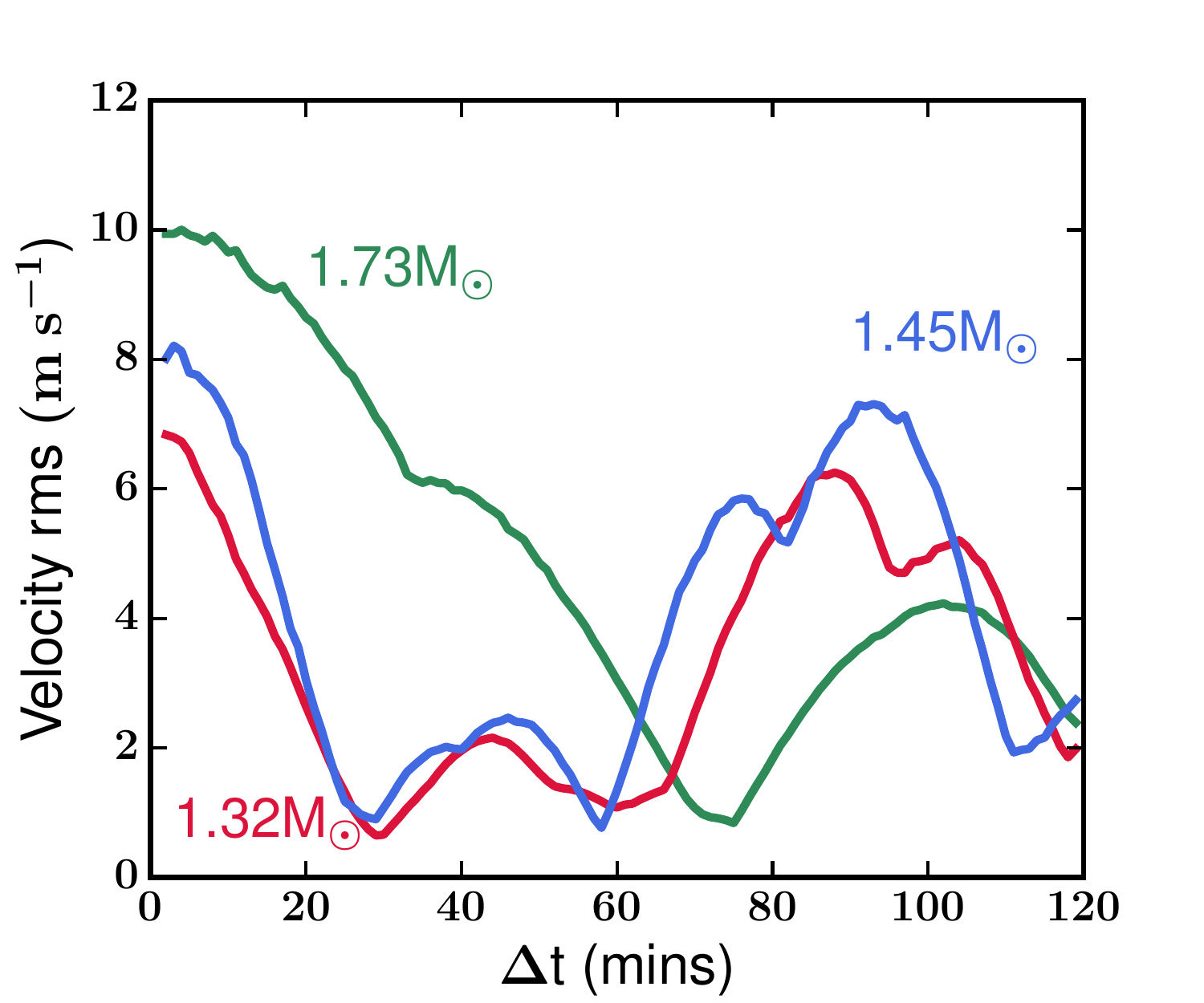}
\caption{RMS scatter of three simulated stars' radial velocity measurements as a function of the time between intra-night observations for a giant star with a mass of 1.73$\msol$ (green curve), a subgiant with a mass of 1.45$\msol$ (blue curve), and a subgiant with a mass of 1.32$\msol$ (red curve).}
\label{fig:sim} 
\end{figure}

\begin{figure}[ht] %%%%%%% FIGURE 
\includegraphics[scale =0.6]{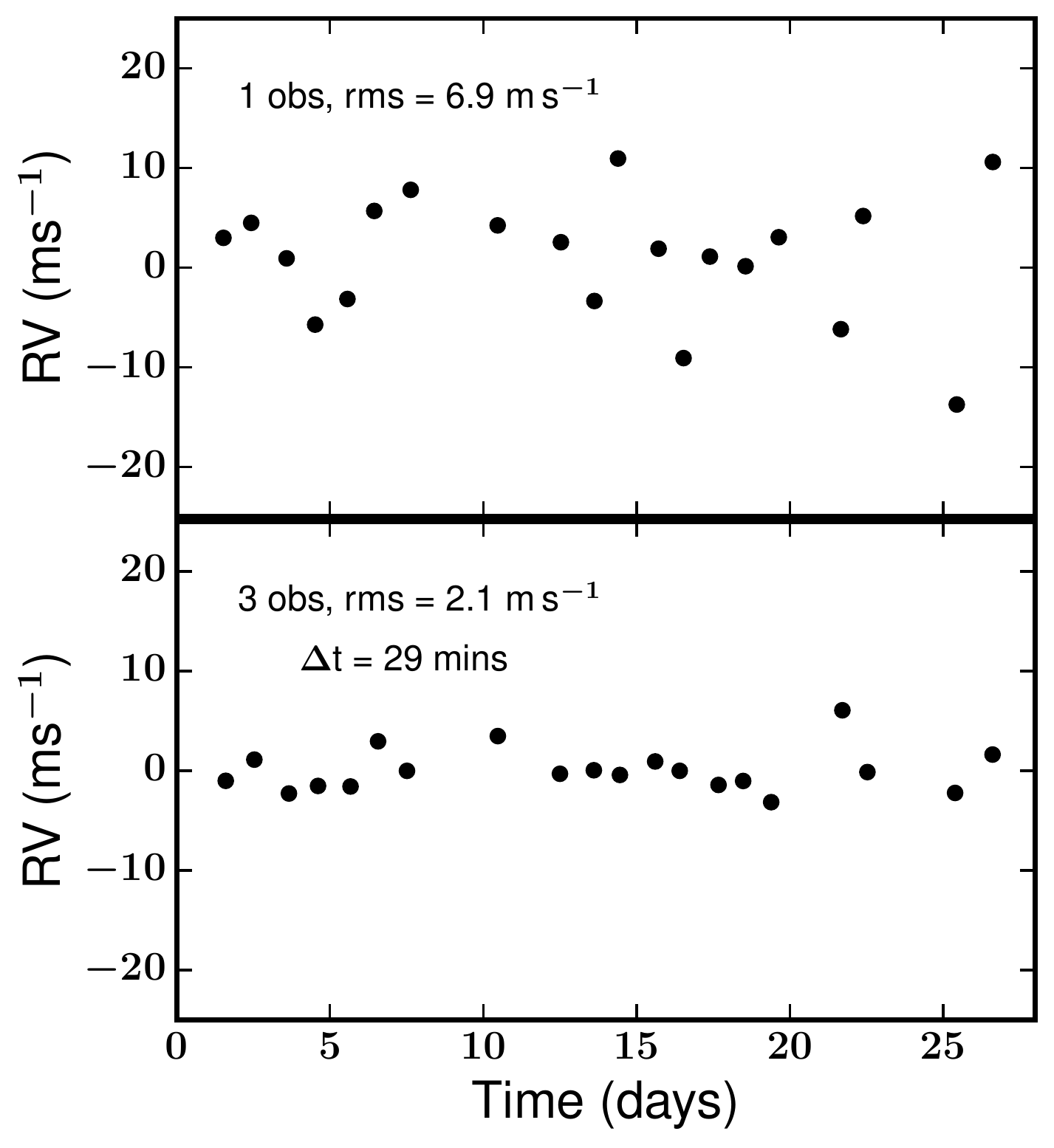}
\caption{Simulated radial velocity time series for data sampled once per night (top) and three times per night (bottom) separated by 29 minutes, the calculated $\Delta t$ for a star of 1.32$\msol$ in Figure \ref{fig:sim}.}
\label{fig:rvs}
\end{figure}

For a given star and $\Delta t$ we used Monte Carlo methods to simulate 1000 individual p-mode RV time series. For each time series, we simulated taking observations over seven consecutive days, with three measurements per night separated by a given $\Delta t$, the time between successive intra-night observations. We averaged the three measurements from a single night and calculated the RMS over the course of seven nights. We then computed the mean RMS of the 1000 trials. We did this for each value of $\Delta t$ (2-120 minutes) and searched for the minimum RMS value as a function of $\Delta t$. The minimum $\Delta t$ determines the observing strategy that maximizes the reduction of stellar noise due to p-mode oscillations. An example of this analysis is shown for a giant star (1.73$\msol$), and two subgiants with masses of 1.45$\msol$ and 1.32$\msol$ in Figure \ref{fig:sim}. Figure \ref{fig:rvs} shows an example of a simulated time series randomly sampled three times per night (top panel) and sampled three times per night at the optimal $\Delta t$ of 29 minutes (bottom panel) determined for the 1.32$\msol$ star in Figure \ref{fig:sim}. The RMS is reduced by a factor of three between the two time series.

\section{Subgiant Data}
\begin{figure}[ht] %%%%%%% FIGURE 
\includegraphics[scale = 0.6]{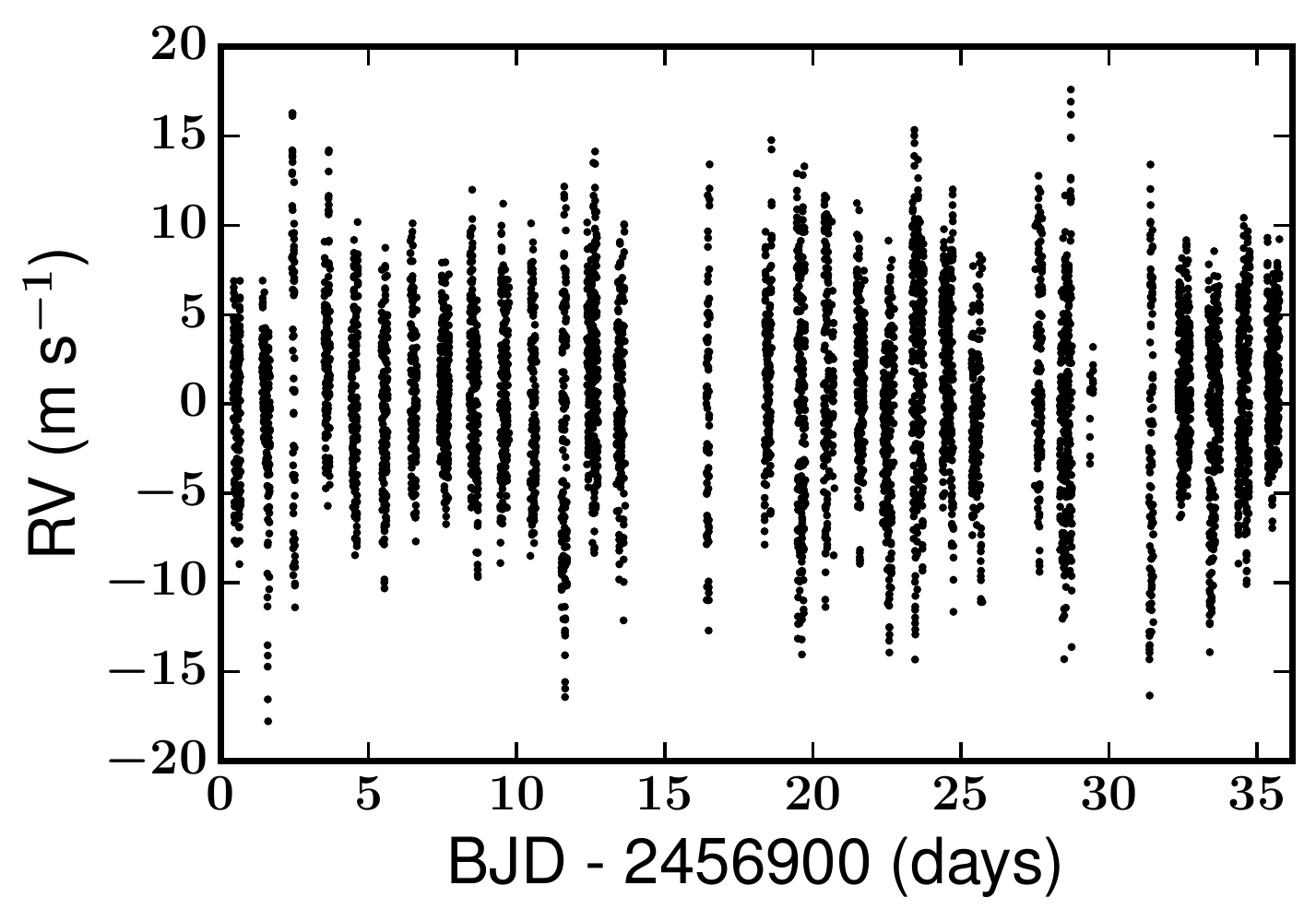}
\caption{Radial velocity data of $\gamma$ Cep from \citep{Stello2017}}
\label{fig:gcdat}
\end{figure}

Our simulated radial velocity time series are based on simplified empirical relationships scaled from the Sun. These relationships, although valid for Sun-like stars, may not be generally applicable to other stars, especially more evolved stars that have larger p-mode amplitudes and longer periods. To test the applicability of our method, we used radial velocity data acquired by \citet{Stello2017} as an ideal test case of radial velocity variations in evolved stars. The radial velocity data were obtained with the echelle spectrograph on the 1 meter Hertzsprung SONG telescope located on Tenerife \citep{Andersen2014}. We obtained 30 nights of radial velocity measurements of the evolved star $\gamma$ Cep from \citet{Stello2017}. On average, each night had 188 observations with 120 second exposures.  Of the stars in the \citet{Stello2017} sample, $\gamma$ Cep was selected because it was the most well sampled over the longest timescale. The data are shown in Figure \ref{fig:gcdat}.

\begin{figure}[ht] %%%%%%% FIGURE 
\includegraphics[scale = 0.6]{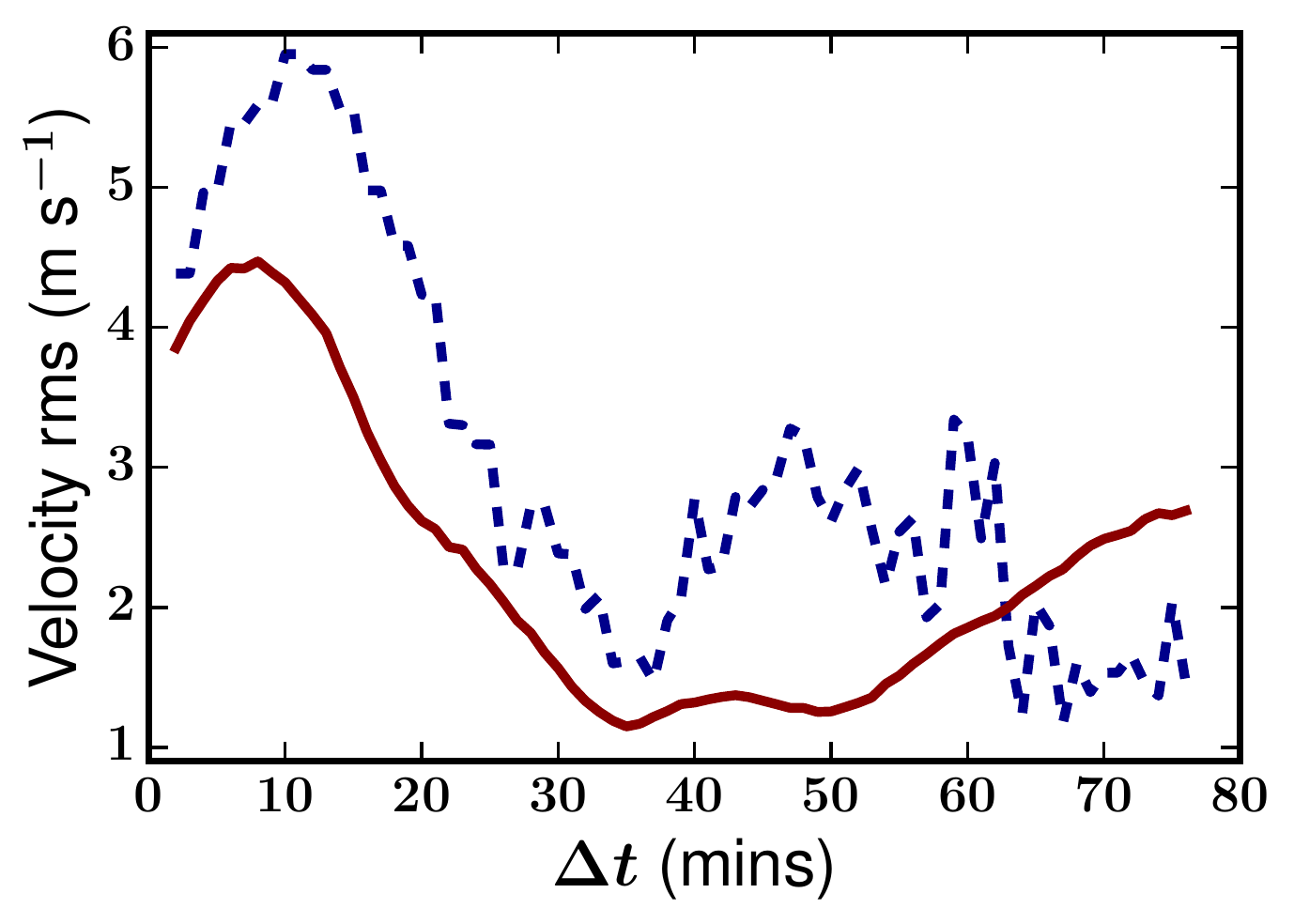}
\caption{RMS scatter of the simulated (solid red line) and actual radial velocity measurements (dashed blue line) as a function of $\Delta t$, the time between intra-night observations}
\label{fig:dtop}
\end{figure}

We applied our methods outlined in Section 2 to calculate the optimal $\Delta t$ for $\gamma$ Cep using stellar properties from \citet{Stello2017} to be 35 minutes. $\gamma$ Cep has a mass of 1.32$\msol$, which is within the mass range we wish to explore, radius of 4.88$\rsol$, luminosity of 11.17$\lsol$, and effective temperature of 4764 K. For comparison,  we also searched for the minimum RMS as a function of $\Delta t$ in the SONG observations which is shown by the dashed line in Figure \ref{fig:dtop}. Our simulations capture the general shape of the RMS as a function of $\Delta t$ and sample the global minimum well. Although our simulations do not treat all of the physical processes of the oscillations, such as mode damping or re-excitation, and make assumptions about the values of mode lifetimes, they appear to reproduce the dominant radial velocity processes that are important for our study. 

In order to test the ability of our method to recover planets located in the Planet Desert, we injected a planetary signal into the observations of $\gamma$ Cep that were obtained by \citep{Stello2017}. We stepped through different values of K, the radial velocity semi-major amplitude, in order to determine what semi-amplitudes would be detectable.  K ranged from (0.1 - 300) m~s$^{-1}$ reflecting planetary masses in the range of (2-3000)M$_{\oplus}$ around a star with mass equal to that of $\gamma$ Cep. For each value of K, 1000 iterations of a planet were simulated as the combination of p-mode oscillations from \citet{Stello2017} and a Keplerian orbit with a period of 10.0 days, zero eccentricity, and a random orbital phase. 

For every iteration of the 1000 that were completed for each K value, each had a randomly chosen start time, and was sampled three times per night separated by 35 minutes, the calculated optimal $\Delta t$ for $\gamma$ Cep. For comparison, the data were also sampled with three consecutive observations per night. This comparison is motivated by the standard procedure used for surveys of evolved stars \citep{Johnson2007} where consecutive observations were used to maximize signal to noise but avoid saturation. In addition to comparing with three consecutive observations per night, we also tested against the best possible outcome of observations: observations where there are no p-modes presents, so the only noise contribution comes from photon noise. 

 We sampled the observations from \citet{Stello2017} according to orbital phase coverage of eight observing days per orbital period. This phase coverage is a guideline and it should be noted that increasing or decreasing the number, $N_{\rm obs}$,  of observations per orbital period will affect the observed RMS by $\sqrt{\rm N_{\rm obs}}$. Noise from a random Gaussian distribution centered at 2 m~s$^{-1}$ was added to each measurement to account for measurement uncertainty. For consistency, each comparison data set (three separated by $\Delta t_{\rm opt}$, three consecutive,  and three consecutive - no p-modes) has the same number of data points, N = 36. 

For each injected planet described above, we then computed a Lomb-Scargle Periodogram to determine the period of each individual data set. We deemed the planet to be recovered if the maximum peak of the LS periodogram was consistent within 1\% of the true period of 10.0 days, i.e. \Big[$\Delta$P$_{\rm obs}$/P$_{\rm true} <$ 0.01\Big].

\begin{figure}[ht] %%%%%%% FIGURE 
\includegraphics[scale =0.6]{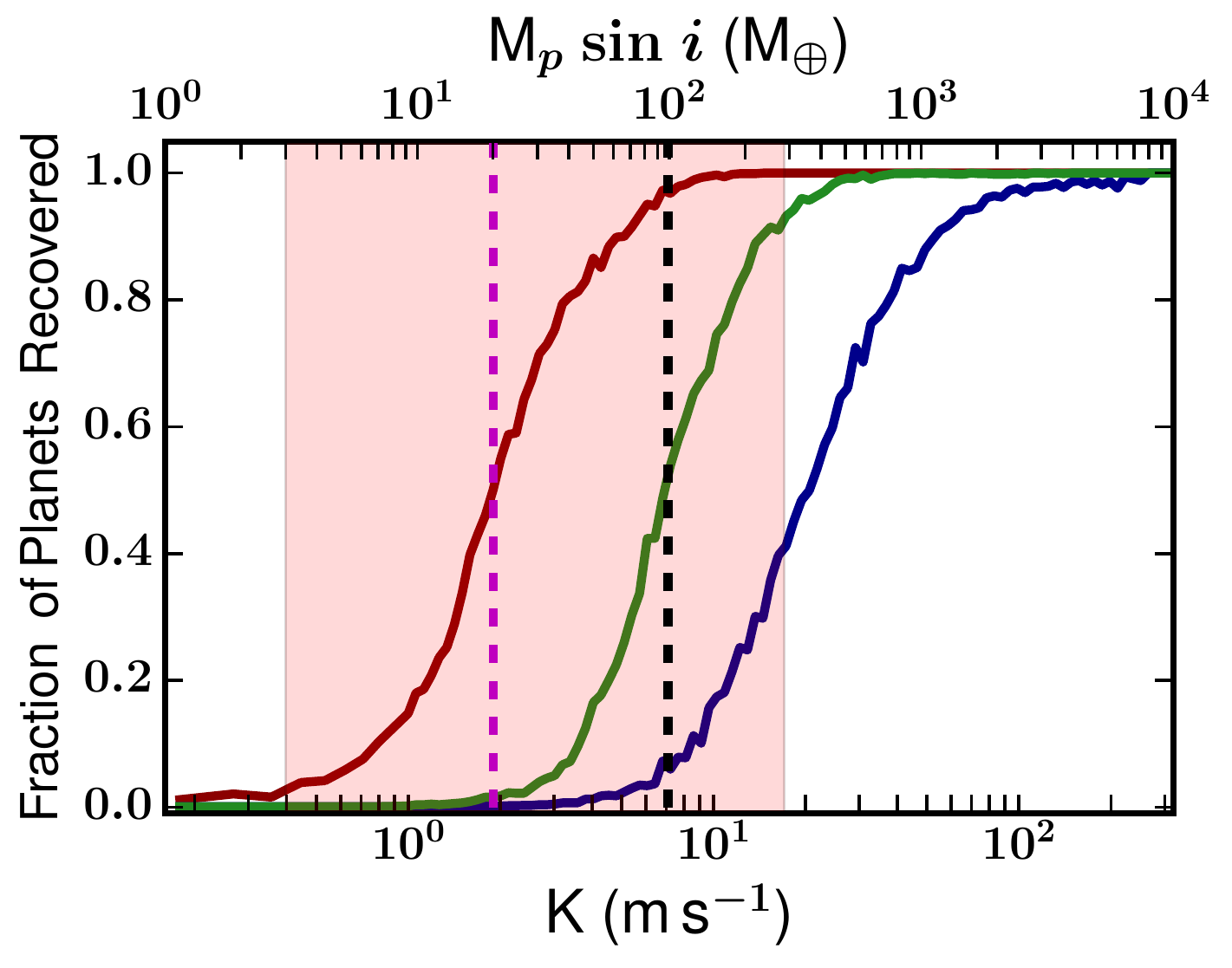}
\caption{Fraction of planets recovered as a function semi-amplitude K. The red curve is the result of taking three consecutive measurements in one night where no p-mode oscillations are present and thus are photon noise limited. This represents the idealized noise floor.  The green curve is the result of taking three measurements per night separated by a calculated optimum $\Delta t$. The blue curve is result of three consecutive measurements in one night. The vertical dashed magenta line represents a Neptune-mass planet around $\gamma$ Cep. The vertical dashed black line represents a Saturn-mass planet.}
\label{fig:trec}
\end{figure}

We show in Figure \ref{fig:trec} the fraction of planets recovered out of 1000 trials as a function of both semi-amplitude, K, and minimum planet mass in Earth masses. Our method recovers more injected planets than the standard observing strategy for subgiant stars. It should be expected that our method does not recover as many planets as the the simulation where no p-modes were present as this represents the idealized noise floor. 

\begin{figure}[ht] %%%%%%% FIGURE 
\includegraphics[scale = 0.6]{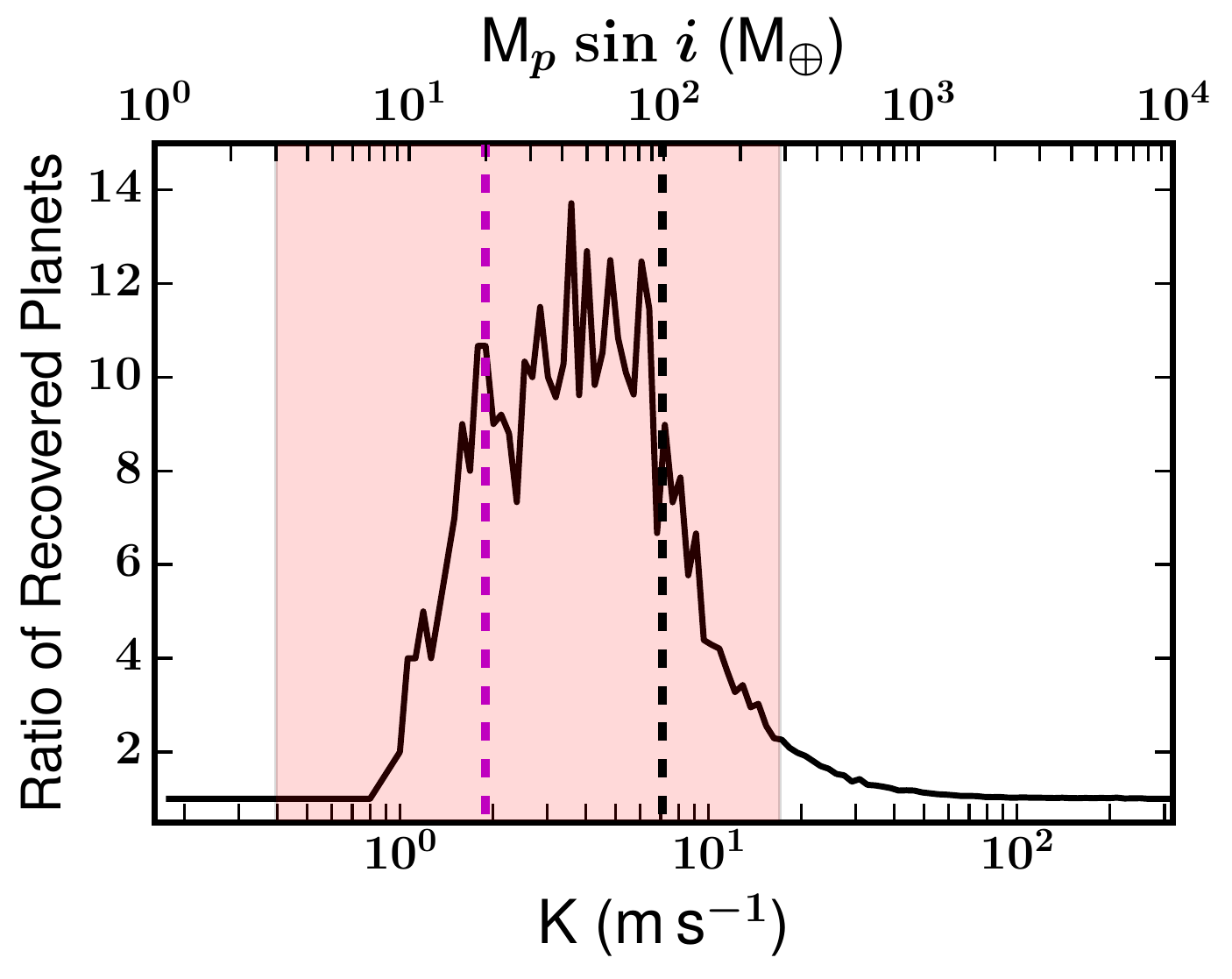}
\caption{The ratio of recovered planets using our method of three observations separated by an optimal $\Delta t$ (green curve in Figure \ref{fig:trec}) to the number of planets recovered using the standard observing procedure for subgiant stars (blue curve in Figure \ref{fig:trec}) as function of the semi-amplitude K  (solid black line). The vertical magenta dashed line represents a Neptune-mass planet around $\gamma$ Cep. The vertical black dashed line represents a Saturn-mass planet. Where the ratio goes to unity for planets that are less massive than a Neptune-mass are instances where the ratio was either zero or undefined by a division by zero. Those instances have been changed to ones for clarity. Note that this does not change our result.}
\label{fig:massdiff}
\end{figure}

Figure \ref{fig:massdiff} is the ratio of the number of planets recovered using our method, three observations optimally sampled (green curve Figure \ref{fig:trec}), to the number of planets recovered using the standard observing strategy for subgiant stars (blue curve Figure \ref{fig:trec}). Our method recovers a factor of ten more planets than the standard observing strategy, specifically in the range of planet masses between Neptune and Saturn shown as the dashed magenta and black lines respectively.

\section{Quantifying $\Delta$t}
In this section we develop a parameterization of the relationship between $\Delta t$, the optimal time between intra-night observations, and the mass and radius of the star. This relationship can be used to inform future observations of subgiant radial velocity surveys focused on planet detection. In the development of this parameterization, we examined the relationships between $\Delta t$ and other stellar parameters such as luminosity and effective temperature, but determined that mass and radius encapsulate the primary contributions to $\Delta t$.

To develop the numerical model, we generated a population of stars with the MESA Isochrones and Stellar Tracks (MIST) \citep{Choi2016} with (1.98 $< \logg <$ 4.08) and (4660~K $<$ T$_{\rm eff}$  $<$ 6130~K) representing stars in the subgiant region of parameter space. We produced stellar masses, luminosities, radii and effective temperature using MIST models for a sample of 403 stars, which we cut to 145 stars to reflect the stellar mass range we are interested in for this study (1.2 $<$ M/M$_{\odot}$ $<$ 2.5). Stars with greater masses have radii that typically fall into the giant category and have been shown to exhibit mixed oscillation modes, which is beyond the scope of this paper. Stars with masses less than 1.2$\msol$ have p-mode oscillation timescales that are feasible to average out via longer integration times. We determined $\Delta t$ for each star via the method outlined in Section 2. Errors on $\Delta t$ were calculated as the RMS of optimal $\Delta t$s for each of the 1000 trials discussed in Section 2.

We chose our functional form based on the analysis in \citet{KB1995} where many of the empirical relationships developed follow power-laws that are rooted in stellar mass, radius, effective temperature, and luminosity. However, upon inspection, we determine mass, ($M = M_{*}/\msol$ ) and radius, $R = R_{*}/\rsol$ have the largest contributions to $\Delta t$. 
We assume that mass and radius have are the primary contributors to $\Delta t$ and thus decided on the form:

\begin{equation}
\label{eq:dt}
\Delta t(M,R) (\rm{mins}) = C \cdot M^{\alpha} R^{\beta}
\end{equation}

\begin{figure}[ht] %%%%%%% FIGURE 
\includegraphics[scale = 0.61]{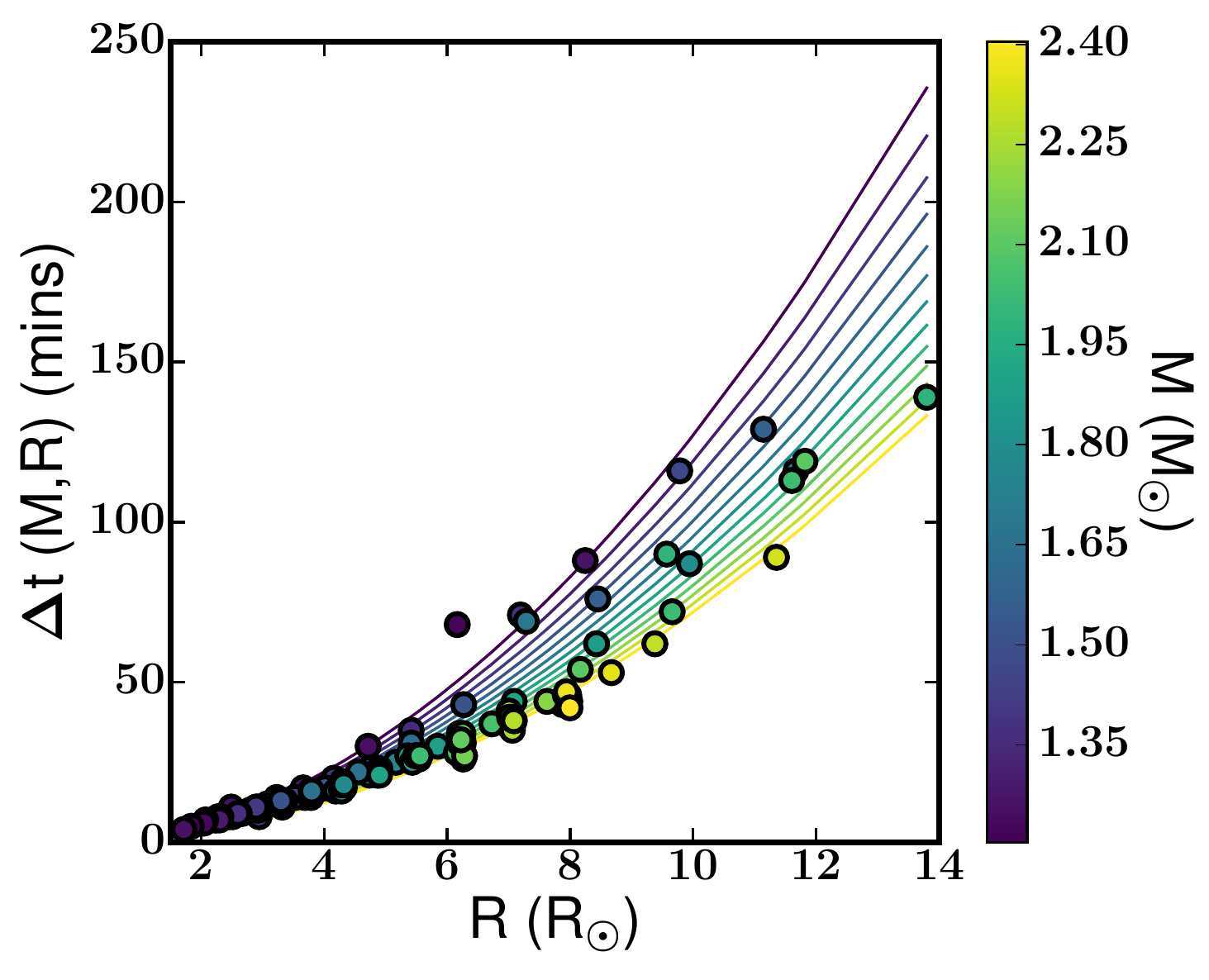}
\caption{$\Delta t$(M,R) in minutes as a function of the stellar Radius, R, for the sample of stars simulated with MIST. The solid lines are $\Delta t$ values calculated using the best fit parameters of our numerical model listed in Table 1. The lines are evaluated as a function of stellar radius at a constant mass [1.20 (purple) - 2.40 (yellow)] denoted by the color of line in the corresponding colorbar. The colorbar is a representation of the stellar mass. Colored points are $\Delta t$ values calculated using methods outlined in Section 2.}
\label{fig:randmfit}
\end{figure}

\begin{figure}[ht] %%%%%%% FIGURE 
\includegraphics[scale = 0.5]{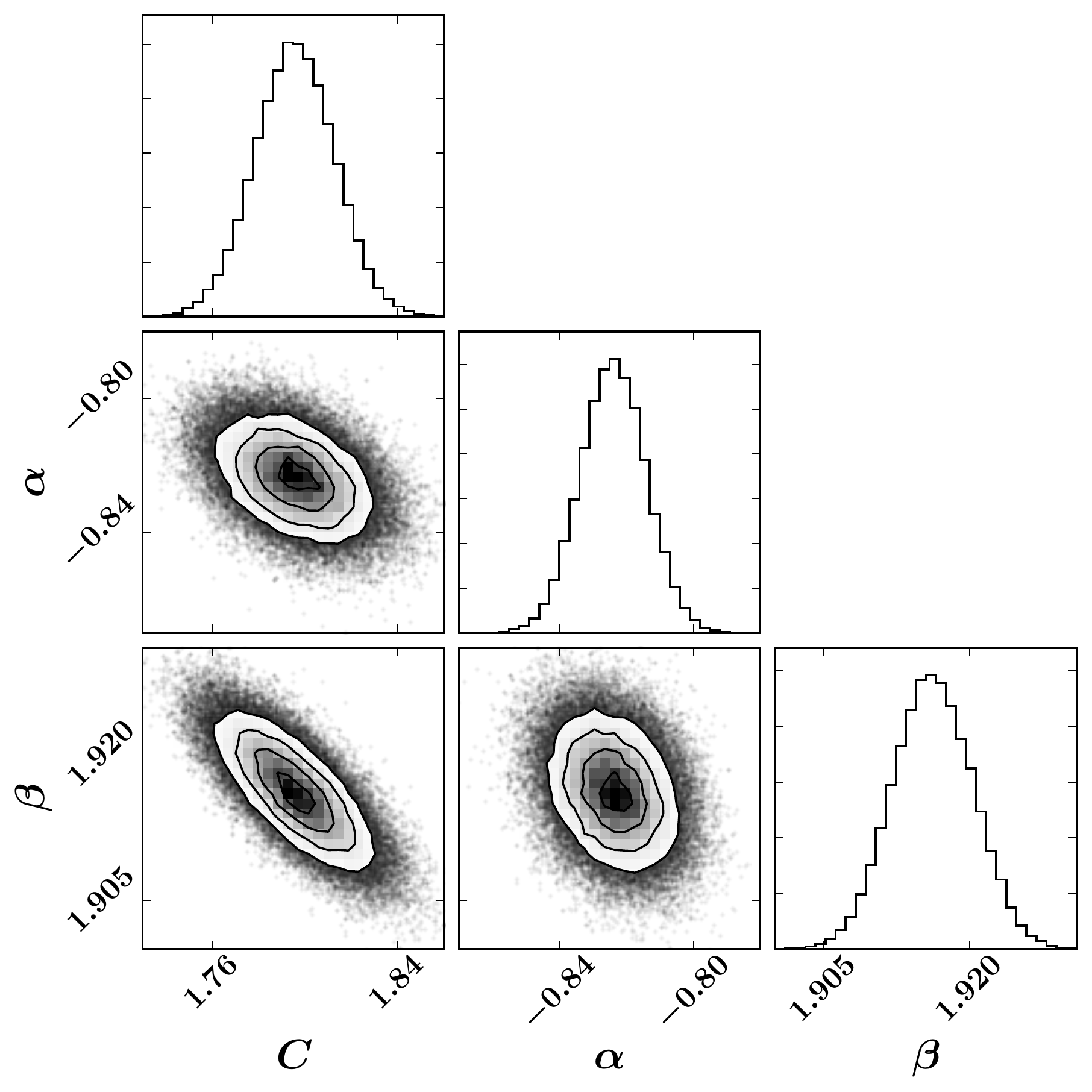}
\caption{Inferred posterior probability distribution for C, $\alpha$, and $\beta$.}
\label{fig:triangle}
\end{figure}

The parameters and their uncertainties were estimated by an MCMC analysis using \textit{emcee} \citep{Foreman2013}.The fit parameters C, $\alpha$, and $\beta$ were evaluated using uniform priors. The best fitting parameters, their uncertainties, and the ranges of the priors are shown in Table 1. The marginalized posterior probability distributions are shown in Figure \ref{fig:triangle} for the parameters of the two dimensional power law we used to model the data. Figure \ref{fig:randmfit} shows a comparison between the fit parameters in Table 1 (solid lines) and the data (colored points) as a function of stellar radius. Each solid line is the optimal observing cadence, $\Delta t$, calculated using the fit parameters in Table 1 as a function of the stellar radius and evaluated at a constant mass indicated by the color of the line in the associated colorbar. The data points trace the lines of $\Delta t$ evaluated at a constant mass well, indicating the numerical model we have chosen to quantify $\Delta t$ as a function of stellar mass and radius is a good representation for the data.

\begin{table}[ht]
\caption{Model Parameters}
\centering
\begin{tabular}{c c c c}
\hline\hline
Parameter & Uniform & Median & 68 \% Confidence\\[0.5ex] 
Name &  Prior & Value & Interval\\ [0.5ex] % inserts table %heading
\hline
$\alpha$&(-1000,1000)&$\alphaval$&($\pm$ 0.010)\\
$\beta$&(-1000,1000)&$\betaval$ &($\pm$ 0.004) \\
C&(-$\infty$,$\infty$)&$\cval$ &($\pm$ 0.019) \\
\hline
\end{tabular}
\label{table:nonlin}
\end{table}

There exists a negative correlation between the contribution to $\Delta t$ from the stellar mass and radius. This correlation makes sense based on the fact that through exploration, we found a very tight relationship between $\logg$, and $\Delta t$. Because $\logg$ is a dependent on both mass an radius, for a fixed $\logg$ if one parameter increases, the other must decrease and vice versa. However, for subgiants, $\logg$ is difficult to measure due to pressure broadening of the spectral lines characteristically used to measure $\logg$. Thus, we decided to use stellar mass and radius as the independent variables from which we model $\Delta t$.

\section{Discussion and Conclusion}
We find that taking three observations per night separated by a calculated optimal $\Delta t$ can reduce the RMS of observations of evolved intermediate mass stars by up to a factor of three over the current radial velocity observing strategies for subgiants: one observation per night or three consecutive observations per night used to maximize signal to noise while preventing saturation. Our method for calculating $\Delta t$ is based on simulating radial velocity variations from scaling relationships that are a function of the stellar mass, luminosity, radius, and effective temperature. We find that these relationships, even though they do not account for a full physical representation of the variations of p-mode oscillations, do a good job of replicating the dominant radial velocity variations as a function of time allowing us to accurately predict the optimal time between intra-night observations that maximizes the reduction of RMS scatter (Figure \ref{fig:trec}).

We tested our method on high cadence radial velocity data of the subgiant $\gamma$ Cep. We find that sampling according to our method where the observations are separated by an optimal $\Delta t$ of 35 minutes results in a higher number of recovered planets than the number recovered taking three consecutive observations per night (Figure \ref{fig:trec}). Shown in Figure \ref{fig:massdiff} is the ratio of the number of trials recovered between three consecutive observations (standard method) and three observations separated by 35 minutes (our method). The peak of this ratio is roughly between a Neptune-like ($\sim$20 M$_{\rm \oplus}$) and Saturn-like ($\sim$100 M$_{\rm \oplus}$ planet around a star with mass equal to $\gamma$ Cep. This population of planets that has been thus far remained elusive in radial velocity surveys of evolved intermediate mass stars. Although in the case of white noise, taking N times as many observations will likely reduce the RMS of the measurements by the $\sqrt{N}$, we have shown the variation can be reduced further due to the nature of p-mode oscillations being correlated noise, and thus we can beat white noise averaging by observing on timescales that are informed by the periods and amplitudes of the p-mode oscillations.

It should be noted that an alternative method for handling the correlated noise of the star is to model it using a Gaussian process (GP). Recent studies by \citet{Brewer2009,Haywood2014,Grunblatt2016,Grunblatt2017,Faria2016,Foreman2017} and \citet{Farr2018} have shown that planets previously hidden by correlated noise from the star, such as granulation or oscillations, can be detected using a model that is the combination of a transit or Doppler planet model, wherein the GP is used to account for the stellar noise. For single mode oscillations specifically the physically motivated GP kernel is a damped exponential, being the impulse response function of a driven, damped harmonic oscillator \citet{Brewer2009}. This incorporation of a GP has proven to be successful at modeling the correlated noise in planet searches and additionally can provide asteroseismic properties, which can ultimately lead to estimates for the evolutionary state of the stars.

Gaussian processes are robust in their ability to handle poorly sampled data and offer an explicit solution to modeling a variety of correlated noise sources in stars. However, GPs are intrinsically computationally expensive, often scaling with the number of data points cubed. \citet{Kelly2014,Foreman2017} have provided GP frameworks that scale linearly with the number of data points making them computationally practical. However this is limited to one-dimensional problems and thus has significant limitations for multi-dimensional problems; in our case for stars that show a comb of oscillation modes as opposed to single modes that have been the focus of previous studies that use GPs. 

Although our method is not based on an explicit solution and requires regularly sampled data, it is computationally inexpensive and provides a clear recipe for planning optimally sampled RV observations for subgiant stars. Ultimately, it should be noted that these two methodologies do not have to be mutually exclusive. An optimal sampling method, like the one we present in this paper, in addition to a GP may provide the best mechanism to find planets located in the Planet Desert, but that exploration is beyond the scope of this paper.

We find $\Delta t$ can be estimated from the stellar mass and radius using the following equation:

\begin{equation}
\label{eq:dt_params}
\Delta t(M,R) (\rm{mins}) = \cval (M^{\alphaval} R^{\betaval})
\end{equation}

Although stellar radius can be determined fairly easily to a high degree of precision using effective temperature and bolometric luminosity, especially with data releases from $Gaia$, the same cannot be said for the determination of stellar mass. However, we find that a mass estimate within 20\% is still able to encapsulate a $\Delta t$ that reduces the observed RMS scatter due to p-mode oscillations. We also find that the contribution from mass to $\Delta t$ can be ignored if one is willing to accept a three minute error in $\Delta t$ translating to a 15\% increase in RV RMS. However, it should be noted that this empirical relationship is to be used primarily for radial velocity surveys where radius and mass determination is fulfilled given the necessary conditions for determining Keplerian orbital parameters. 

The basis of our proposed method, in principle, could be applied to all stars exhibiting p-mode oscillations. However, for stars that are more evolved on the giant branch our methods break down due to the nature of mixed modes that act as p-mode oscillations in the outer convective zone and g-modes, where gravity is the restoring force, in the interior of the star. This mixed mode behavior shifts the observed frequencies from where pure p- or g-modes would occur making it difficult to predict the timescales of the observed oscillations and thus is beyond the scope of this paper.

Surveys of evolved intermediate mass stars uncovered a rich population of Jupiter mass planets but show a paucity of close-in ($a < 0.6$ AU), intermediate period (5 $< P <$ 100 days), low-mass ($M_{\rm planet} <$ 0.7$M_{\rm Jup} $), known as the `Planet Desert'. The discovery of an abundance of Jupiter mass planets coupled with a lack low-mass planets, might be taken as evidence that massive stars favor the production of massive planets. This interpretation is not necessarily the case given the fact that asterometric noise could be hiding less massive planet. In recent studies of planet occurrence of Kepler stars \citep{Howard2012,Fressin2013,Petigura2013,Mulders2015,Fulton2017} studies suggest that there is an increase in planet occurrence with decreasing planet size. Interestingly, \citet{Fulton2017} find that in particular planets with radii 2-4R$_\odot$ at periods less than 100 days have an occurrence rate of 36 \%. This suggests that there could exist a large population of low-mass, close-in planets around evolved IM stars.

The 5-10 m~s$^{-1}$ jitter levels of evolved intermediate mass stars could be concealing this population of close-in low-mass planets. In this paper, we have demonstrated that the implementation of our observing strategy based upon the star's intrinsic stellar properties could result in the discovery of planets located in the Planet Desert. These results aim to inform future radial velocity surveys of evolved intermediate mass stars, like those aimed at following up objects of interested from the Transiting Exoplanet Survey Satellite (TESS).  \citet{Barclay2018} predict that with two minute cadence TESS will find $\sim$300 close-in ($P < 85$ days) planets with R$_p > 4$R$_{\oplus}$. Of those stars, \citet{Barclay2018} estimate 25\% will be around evolved IM stars. These stars were serendipitously included in the TESS Input Catalog Candidate Target List because they tend to impersonate main sequence stars based on their magnitudes and pre-$Gaia$ proper motions \citep{Stassun2017,Barclay2018}. High-cadence photometric observations of these bright evolved IM stars provide a means for transiting exoplanet detections in the Planet Desert and information about the asterosiesmic properties of these stars making them interesting targets for a broad astronomical community. RV follow-up of these transit signals will be key for understanding the distribution of planet properties around IM stars. Our method provides a guide for planning this type of RV follow-up observations. Determining the existence of these planets would aid the understanding of planetary formation and demographics both around intermediate mass stars and more generally as a function of stellar mass.

\acknowledgments{The authors thank the anonymous referee for their helpful comments and suggestions to this manuscript. The authors also greatly appreciate Dennis Stello for providing access to SONG data made with the Hertzsprung SONG telescope operated on the Spanish Observatorio del Teide on the island of Tenerife by the Aarhus and Copenhagen Universities and by the Instituto de Astrofísica de Canarias. A.A.M would like to thank R.D Haywood for helpful discussions on the topic of Gaussian processes. A.A.M is supported by NSF Graduate Research Fellowship, Grant No. DGE1745303. J.A.J is grateful for the generous grant support by the David and Lucile Packard foundation. Work performed by P.A.C. was supported by NASA grant NNX13AI46G.} 

%\bibliographystyle{apj}
%\bibliography{references}

\end{document}